\documentclass{aa}
\usepackage{graphicx}
\begin{document}
\title{W Hya : molecular inventory by ISO-SWS
\thanks {Based on observations with ISO, an ESA project with
 instruments funded by ESA Member States (especially the PI
 countries:
 France, Germany, the Netherlands and the United Kingdom) with the
 participation of ISAS and NASA.} }

\author{K. Justtanont \inst{1} \and
        T. de Jong\inst{2,3} \and
        A.G.G.M. Tielens\inst{4,5} \and
        H. Feuchtgruber\inst{6} \and
        L.B.F.M. Waters\inst{3}
        }

\offprints{K. Justtanont}

\institute{Stockholm Observatory, AlbaNova, Dept. of Astronomy, SE-106 91 Stockholm, Sweden
\and SRON-Utrecht, Sorbonnelaan 2, NL-3584 CA Utrecht, The Netherlands
\and University of Amsterdam, Kruislaan 403, NL-1098 SJ Amsterdam,
The Netherlands
\and Kapteyn Institute, P.O. Box 800, NL-9700 AV Groningen, The
Netherlands
\and SRON-Groningen, P.O. Box 800, NL-9700 AV Groningen, The Netherlands
\and Max-Planck Institut fuer extraterrestrische Physik, Giessenbachstrasse, 
D-85740 Garching, Germany
}

\date{Received ; accepted }

\abstract{
Infrared spectroscopy is a powerful tool to probe the inventory of solid
state and molecular species in circumstellar ejecta.  Here we analyse
the infrared spectrum of the Asymptotic Giant Branch star W Hya,
obtained by the Short and Long Wavelength Spectrometers on
board of the Infrared Satellite Observatory.  These spectra show evidence
for the presence of amorphous silicates, aluminum oxide, and
magnesium-iron oxide grains.  We have modelled the spectral energy
distribution using laboratory measured optical properties of these
compounds and derive a total dust mass loss rate of $3\times 10^{-10}$
M$_{\odot}$ yr$^{-1}$.  We find no satisfactory fit to the 13$\mu$m 
dust emission feature and the identification of its carrier is still
an open issue.
We have also modelled the molecular absorption bands due to H$_{2}$O, OH,
CO, CO$_{2}$, SiO, and SO$_{2}$ and estimated the excitation temperatures 
for different bands which range from 300 to 3\,000K.
It is clear that different molecules giving rise to these absorption
bands originate from different gas layers.
We present and
analyse high resolution Fabry-Perot spectra of the three CO$_{2}$ bands in
the 15$\mu$m region.  In these data, the bands are resolved into individual
Q-lines in emission, which allows the direct determination of the excitation
temperature and column density of the emitting gas.  This reveals the
presence of a warm ($\simeq$ 450K) extended layer of CO$_{2}$,
somewhere between the photosphere and the dust formation zone. The gas
in this layer is cooler than the 1\,000K CO$_{2}$ gas responsible for the
low-resolution absorption bands at 4.25 and 15$\mu$m. The rotational and
vibrational excitation temperatures derived from the individual Q-branch
lines of CO$_{2}$ are different ($\sim$ 450K and 150K, respectively) so that
the CO$_{2}$ level population is not in LTE.
}
\maketitle

\section{Introduction}

One of the characteristics of Asymptotic Giant Branch (AGB) stars
is their mass loss. As a star evolves up the AGB, its mass loss rate
increases. Just before it evolves off to become a planetary nebula,
it sheds most of its mass in a short time, i.e., superwind. Signatures
of mass loss can be studied in the infrared by tracing the dust 
(e.g., Bedijn \cite{bedijn}, 
Justtanont \& Tielens \cite{jt92}), 
and in the sub-millimetre by using molecular rotational lines such as
CO (e.g., Knapp \& Morris \cite{kn85}, Groenewegen \cite{groen},
Skinner et al. \cite{cjs}). 

An example of an early AGB star is W Hya. It has a relatively
low mass loss rate as deduced from the weak CO millimeter observations
(Wannier \& Sahai \cite{wannier}; Bujarrabal et al. \cite{bujar};
Cernicharo et al. \cite{cerni}). It is classified as a semiregular
variable (SRa) from its light curve. The visual magnitude ranges
from 6 to 10 over a period of 382 days (Lebzelter et al. 
\cite{leb}). It exhibits SiO
(e.g., Schwartz et al. \cite{schwartz}; Clark et al. \cite{clark}),
H$_{2}$O (e.g., Spencer et al. \cite{spencer}; Lane et al. \cite{lane}) and
OH main line masers (e.g., Szymczak et al. \cite{szym};
Chapman et al. \cite{chapman}). It has a weak silicate dust emission at
10$\mu$m along with a smaller emission feature at 13$\mu$m. The carrier
of the latter is still being debated. W Hya is one of the first AGB stars 
for which H$_{2}$O rotational emission was reported after the launch of the
Infrared Space Observatory (ISO, Kessler et al. \cite{kessler}).
Neufeld et al. (\cite{neufeld}) and Barlow et al. (\cite{barlow})
reported that these lines were detected in Short and Long Wavelength
Spectrometers (SWS and LWS), respectively. Both
authors presented models accounting for the emission. Recently,
Zubko \& Zuckerman (\cite{zubko}) also reported a model predicting
H$_{2}$O line fluxes observed by ISO. 
Also, W Hya has been detected in the ortho ground state
line by SWAS at 557GHz (Harwit \& Bergin \cite{harwit})
In our paper, we mainly
discuss the SWS spectra taken during the ISO mission. We also present
the SWS Fabry-Perot spectrum of W Hya centered on CO$_{2}$ bands.
This band has been reported by Ryde (\cite{ryde}) and Justtanont
et al. (\cite{jus98}).

\section{Observations}

Spectroscopic observations of W Hya were taken aboard ISO
using both the SWS (de Graauw et al. \cite{thijs}) and LWS
(Clegg et al. \cite{clegg}). The SWS data were taken using the standard
AOT1 (full scan from 2.38-45.2$\mu$m) speed 3, giving an
average resolution of 510. We also obtained the spectrum
using the full grating resolution of 2000 between 12-16$\mu$m
(AOT6) in order to study the CO$_{2}$ bands. The data were reduced 
using the data reduction package provided by the SWS calibration
team based at SRON-Groningen. We were able to observe the three
CO$_{2}$ bands using the SWS Fabry-Perot (FP) 
centering on the three Q-branches in order to resolve individual transitions.
Care was taken to deglitch the spectrum due to cosmic ray hits on
the detectors at the time of observation. As the data were taken in
small wavelength intervals, each subband spectrum had to be stitched 
together. The final spectra were rebinned to a resolution of 4\,10$^{4}$,
with an oversampling of 6. Also the flux level for 
each wavelength band has been normalised to the AOT6 flux.

The LWS data were taken from the archive (PI: M. Barlow).
The reduction was done using the ISAP data reduction package
provided by IPAC.

\section{Dust and gas mass loss rates}

In this section, we present a model fit to the infrared spectral energy
distribution (SED) in order to derive the dust mass loss rate. This
is then used to calculate the rate of gas which is driven by the dust to 
the observed terminal velocity by momentum transfer between gas and dust. 
At the same time, the temperature structure of the
circumstellar envelope is calculated. The main gas heating is by the
dust drag while cooling is done through adiabatic expansion of the gas and
radiative cooling via rotational molecular line emission. We
also include vibrational cooling by H$_{2}$ 
(Goldreich \& Scoville \cite{gold}, Justtanont et al. \cite{jus94}).
Hence the dust and gas mass loss rates are derived in a 
self-consistent manner. The issue of mass loss rates is readdressed
here as there is a wide range of reported rates from 3\,10$^{-5}$
(Neufeld et al. \cite{neufeld}) to 5.2\,10$^{-8}$ M$_{\odot}$ yr$^{-1}$ 
(Wannier \& Sahai \cite{wannier}); see also Zubko \& Elitzur (\cite{zubko})
for a recent reassessment of this controversy.

\subsection{The dusty envelope}

To derive the dust mass loss rate, we have fitted the SED
of W Hya using ISO SWS and LWS data, together
with photometry from ground based observations (Wilson et al. \cite{wilson})
and the IRAS point source fluxes. We adjusted the flux levels for
the LWS spectrum to coincide with the IRAS 60$\mu$m flux which also
joins the SWS spectrum very well. There are broad features present in the
LWS spectrum which are due to the internal reflection within the instrument
at 54 and 110$\mu$m. Here, we calculate the radiative
transfer through the dust shell, using a code based on Haisch (\cite{haisch})
which takes into account thermal emission and reemission and 
multiple scattering. It also allows for a grain size distribution
which we assumed to be the same as the interstellar grain sizes
(Mathis et al. \cite{mrn}, hereafter MRN). We also employ multiple grain species.

The input parameters are listed in Table~\ref{tab:sed}. The distance
is taken from the Hipparcos catalogue (ESA \cite{hip}). 
From the study by Cami (\cite{jan}), the emission features between 10-20$\mu$m
are a combination of amorphous silicates giving rise to the 10$\mu$m
emission, compact Al$_{2}$O$_{3}$ which gives a feature at 11$\mu$m,
spinel (MgAl$_{2}$O$_{4}$) is responsible for the 13$\mu$m feature 
(Posch et al. \cite{posch}), and the 19$\mu$m emission which is attributed
to Mg$_{0.1}$Fe$_{0.9}$O. In our attempt to fit these dust features,
we used silicate dust properties employed by Justtanont \& Tielens 
(\cite{jt92}). For other compounds,
we extracted the optical constants from the University of Jena database
and calculate the extinction coefficients using the Mie theory.

\begin{figure*}
\centering
\includegraphics[width=17cm]{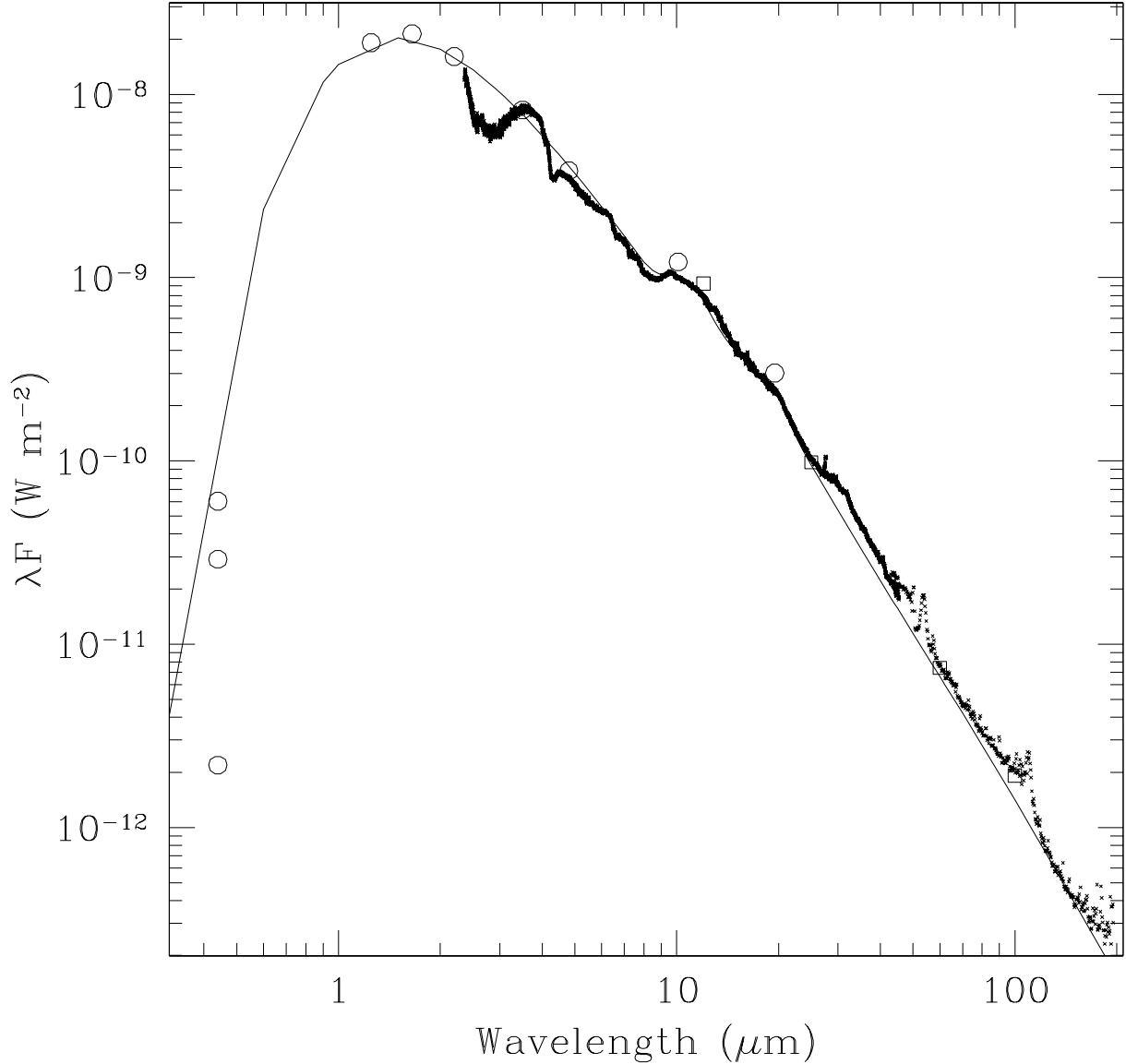}
\caption{A fit to the spectral energy distribution of W Hya
(solid line) with a three-dust component of amorphous silicates,
Al$_{2}$O$_{3}$ and MgFeO. Open circles are from simbad (V magnitudes)
and ground based observations (Wilson et al. \cite{wilson}) 
while filled squares are IRAS PSC and crosses are SWS and LWS data.
Note that the model does not account for the 13$\mu$m emission feature.}
\label{fig:sed}
\end{figure*}

\begin{figure}
\resizebox{\hsize}{!}{\includegraphics{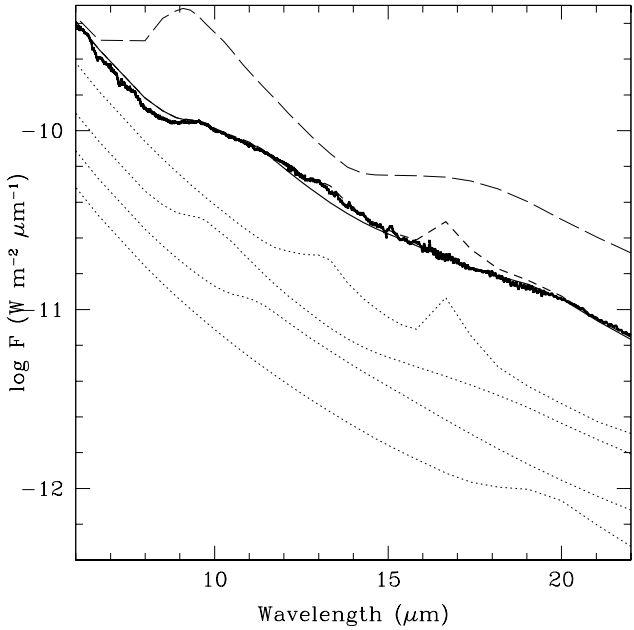}}
\caption{The plot of a close-up of the dust features. 
The long dashed line represent a model with Zubko \& Elitzur (\cite{zubko})
parameters. The heavy solid line is the three dust model (Fig.~\ref{fig:sed})
while the heavy dashed line is a model which includes spinel.
The contribution from individual dust component is shown as dotted
lines. From top to bottom : spinel, amorphous silicate, Al$_{2}$O$_{3}$,
and Mg$_{0.1}$Fe$_{0.9}$O.}
\label{fig:sed_10um}
\end{figure}

The fit to the SED for a three-dust component model is reasonably good 
for the complete SWS and LWS data coverage (Fig.~\ref{fig:sed}).
The fit is very satisfactory for all observed 
emission features, providing support for the presence of amorphous 
silicates, Al$_{2}$O$_{3}$ and MgFeO, except for the 13$\mu$m feature.  
A model with spinel which produces a fit to the 13$\mu$m
emission feature predicts an additional feature at 16$\mu$m, not seen
in the observation (Fig.~\ref{fig:sed_10um}). Hence, while
spinel provides a good fit to the 13$\mu$m band, this identification
is suspect.  Further laboratory studies on related materials are required 
to test this suggestion.
Onaka et al. (\cite{onaka}) suggested aluminum oxide as a carrier for the
13$\mu$m dust emission and following that,
Kozasa \& Sogawa (\cite{kozasa}) were able to explain the 13$\mu$m feature
using crystalline aluminum oxide core-silicate mantle grains. However, the
optical constants used for aluminum oxide were coarse (1$\mu$m
spacing, Toon et al. \cite{toon}) and the interpolation
of them close to the resonance could result in a misleading profile.
A possible carrier proposed by Speck (\cite{speck}) is
SiO$_{2}$ which has features at 9.5, 13 and 20$\mu$m
while Begemann et al. (\cite{begem}) suggested a silicate-based 
material. For the moment, the issue of the dust species responsible for 
the observed 13$\mu$m feature is still unresolved.

At a distance of 115pc, we derive a total dust mass loss rate of 
$3\,10^{-10}$ M$_{\odot}$ yr$^{-1}$ for the three species of
silicates, Al$_{2}$O$_{3}$ and MgFeO,  assuming a constant 
outflow velocity of 6 km s$^{-1}$ (Bujarrabal et al. \cite{bujar}).
We note here that we assume that the dust and gas velocity is the
same as the fitting of the SED allows us to derive 
$\dot{M}_{d}/v_{\rm dust}$. We will discuss the effect of the dust drift
velocity in Sect. 3.2.
Our input values are very similar to those
used by Zubko \& Elitzur (\cite{zubko}) but we derive a dust mass
loss rate which is lower by a factor of ten.
We investigated whether this could be due to the difference in the dust
properties used. Zubko \& Elitzur used the silicate optical constants
derived for AGN dust (Laor \& Draine \cite{laor}), 
which has its infrared characteristics 
based upon astronomical silicates (Draine \& Lee \cite{draine}).
We have also analysed the observed spectra using dust parameters
for the astronomical silicates
and for Mg$_{0.5}$Fe$_{0.5}$SiO$_{3}$ from the University of Jena database.
The derived dust mass loss rates which fit the observed strength of the
10$\mu$m dust feature agree with our value. 
Hence the difference in
the derived dust mass loss rate is not due to the differences in the
optical properties used. Rather, 
Zubko \& Elitzur (\cite{zubko}) did not attempt to fit the
observed 10$\mu$m feature in detail, but using only ground based-photometry 
and the ISO-LWS spectrum. We emphasize that the high mass
loss rate of Zubko \& Elitzur (\cite{zubko}) gives too strong 
10 and 20$\mu$m silicate emission as shown in
Fig.~\ref{fig:sed_10um}.
We find that extending the calculation to include the dust envelope
size observed by IRAS at 60 and 100$\mu$m (Hawkins \cite{hawkins})
does not affect the far-IR flux as the density is far too low in the
outer part to affect the SED. For our case,
we fitted the silicate 10$\mu$m feature and the other dust species
(except for the 13$\mu$m emission feature) which givess a more reliable
mass loss estimate as dust mass loss rate is proportional to the
strength of the 10$\mu$m silicate feature. However, since we did not
fit the 13$\mu$m feature, our estimate of the mass loss rate should be
viewed as a lower limit, pending the correct identification of the
feature.

If we assume that the elemental abundance of the outflow is that
of the Sun, our estimated dust mass loss rate for each dust 
component indicates that aluminum condenses out at
a higher rate than the silicate, magnesium and iron : the latter
three having abundances an order of magnitude higher than that
of Al (Anders \& Grevesse \cite{anders}). The amount of the silicate 
dust is the same as the aluminum dust while that of MgFeO dust is five times
less (Table~\ref{tab:sed}). Thermodynamic consideration of the
dust condensation sequence starts with gas phase condensation
of Al$_{2}$O$_{3}$, leading finally to Mg- and Fe-incorporation
into silicate grains (Tielens \cite{xander}). Alternatively,
Al and Si condense out at the same time but as Al has a high affinity for
oxygen, aluminum oxide will be formed at a higher
rate than silicate (Stencel et al. \cite{stencel}) thereby
giving rise to an 11$\mu$m feature which is as strong as the
silicate feature even though the abundance of Al is much lower
than that of Si. In either case, the high fraction of Al in dust as compared to Si,
reflects a freeze out of the condensation process well before all the
silicon has been consumed (e.g., well before magnesium-iron silicates condense
out).  W Hya is prototypical for a set of O-rich stars which have 10-20$\mu$m 
spectra which are not dominated by the well-known broad
(magnesium/iron) silicate emission features but rather by a set of oxide
features (Cami \cite{jan}).  We surmise that this difference reflects a steep
density gradient in the dust condensation layer in these sources as
compared to typical Mira variables.  The presence of such a steep
density gradient is also indicated by the asymmetric light curve of
these types of stars.

\subsection{The molecular envelope}

Using the total dust mass loss rate obtained by fitting the SED which drives 
the gas to the terminal velocity, we obtain a dust-to-gas mass ratio of 
8.7\,10$^{-3}$, giving a gas mass loss rate of 3.5\,10$^{-8}$
M$_{\odot}$ yr$^{-1}$ (Model 1)
from solving the equation of motion, assuming that
the gas and dust are fully momentum coupled. 
\begin{equation}
v \frac{dv}{dr} = (\Gamma - 1)\frac{GM_{*}}{r^{2}}
\end{equation}
where $v$ is the gas velocity, G is the gravitational constant, $M_{*}$ 
is the mass of the star, and $\Gamma$ is the ratio of the radiation pressure
force on dust grains to the gravitational force. Following Justtanont 
et al. (\cite{jus94}), $\Gamma$ can be written to include the dust size
distribution, $n_{d}(a)$, and $Q(a,\lambda)$, the radiation pressure efficiency 
for a grain size, $a$, at wavelength $\lambda$ as
\begin{equation}
 \Gamma = \int\int \frac{\sigma_{d} Q(a,\lambda) L(\lambda)}{4\pi c G M_{*}}
           \frac{n_{d}(a)}{\rho} d\lambda~~da
\end{equation}
where $\sigma_{d}$ is the geometric cross section of the dust, $L(\lambda)$
is the luminosity at the wavelength $\lambda$, $\rho$ is the gas density
(see Justtanont et al. (\cite{jus94}) for full details). 
We use radiation pressure for purely silicate dust in the calculation of the
equation of motion. As a matter of fact, we find that the dust velocity 
is much greater than that of the gas due to the drift velocity in stars 
with such a low mass loss rate.
\begin{equation}
 v_{\rm drift}^{2}(a) = \int \frac{Q(a,\lambda)L(\lambda)v}{\dot{M}c} da
\end{equation}
The drift velocity is calculated for each grain size and then weighted
by the size distribution of $a^{-3.5}$. The ratio for the drift velocity for the
largest and smallest grains (0.25 - 0.005$\mu$m) is 8.1. In our calculation, the
heating is dominated by the largest grains (heating $\propto v_{d}^{3}$).
This results in a high heating rate, therefore the calculated gas 
kinetic temperature is very high. In the fitting of the SED, we derived an
$\dot{M}/v_{\rm dust}$ in which we can ignore the drift velocity.
However, if we take the drift velocity into account when calculating the dust 
mass loss rate and then solve for both the equations of motion and energy balance 
in a self consistent way, we derive a mass loss rate of 
1.5\,10$^{-7}$ M$_{\odot}$ yr$^{-1}$ (Model 2) and 
the calculated CO line fluxes are higher
than those observed for the low J transitions (see Table~\ref{tab:tmb}).
From running a number of models with various different densities and 
temperatures, we find that the line flux 
ratio of the CO 2-1/1-0 is a sensitive measure of the
assumed mass loss rate while the temperature plays a minor role
For the observed ratio of 14.7, an upper limit for the mass loss rate is
8\,10$^{-8}$ M$_{\odot}$ yr$^{-1}$ (Model 3, with the temperature derived from
solving the energy balance equation). This model gives the best description
of the observed line fluxes from low to higher transitions.
As a test, we calculated CO line fluxes 
using this mass loss rate and a simple temperature
law $T \propto (r/r_{i})^{-\alpha}$ where $\alpha$ ranges from 0.5 to 0.8.
The resulting line fluxes are insensitive to the change of temperature at
this low mass loss rate ($\leq$ 10\%). This mass loss rate
is within a factor of two of our detailed 
calculation and puts a stringent upper limit on the mass loss rate for W Hya.
We also calculated the expected CO line fluxes using parameters
from Zubko \& Elitzur (\cite{zubko}), including their temperature structure.
The calculated line fluxes are higher than those by Barlow et al. 
(\cite{barlow}), reflecting the higher mass loss rate.

\begin{table}
\caption{Input parameters for modelling the dust and
gas mass loss rates in W Hya}
\begin{tabular}{ll}
\hline 
Stellar mass                  & 1 M$_{\odot}$\\
Stellar effective temperature & 2500 K\\
Stellar radius                & $4\,10^{13}$ cm\\
Outflow velocity              & 6 km s$^{-1}$\\
Distance                      & 115 pc\\
Dust condensation radius      & $2\,10^{14}$ cm\\
Dust mass loss rates      & $1.5\,10^{-10}$ M$_{\odot}$ yr$^{-1}$ - silicates\\
      & 1.3\,10$^{-10}$ M$_{\odot}$ yr$^{-1}$ - Al$_{2}$O$_{3}$\\
      & 2.5\,10$^{-11}$ M$_{\odot}$ yr$^{-1}$ - MgFeO\\
Dust-to-gas mass ratio        & $8.7\,10^{-3}$\\
Gas mass loss rate            & $3.5\,10^{-8}$ M$_{\odot}$ yr$^{-1}$\\
\hline
\label{tab:sed}
\end{tabular}
\end{table}

We calculate the peak main beam antenna temperatures for
Model 1 to be 0.1K for CO J=1-0 observed
with the IRAM telescope (cf. 0.3K from 
Bujarrabal et al. \cite{bujar}) and 0.7K for J=2-1 observed 
with a clear double-peaked profile 
(cf. a flat-top profile with a peak temperature of 2.3K
from Cernicharo et al. (\cite{cerni})),
with a canonical value of the CO abundance of 3\,10$^{-4}$.
Our calculated gas kinetic
temperature, obtained from solving the energy balance,
in the outer part of the shell is very 
high due to the large drift velocity, hence the populations
of low-lying levels are excited to higher levels.
This, coupled with a low density,
results in very low  antenna temperatures for J=1-0 and 2-1 lines. 
In order to reconcile these with the observations, we need to
increase the mass loss rate to 8\,10$^{-8}$ M$_{\odot}$ yr$^{-1}$
(Model 3).
We also calculated expected line fluxes using input
parameters from Barlow et al. (\cite{barlow}) and their temperature,
and compared them to the IRAM results plus J=16-15 and J=17-16 in 
their ISO-LWS spectrum. 
The results of these calculations are summarized in Table~\ref{tab:tmb}. 
Our model predicts line fluxes which are lower than the observed
values, which is due to the low derived mass loss rate for lower transitions
while Barlow et al.'s parameters predict too high line fluxes by a factor of ten.
However, we find that the high
rotational lines are insensitive to the input mass loss rate.
The fact that there is a combination of different
types of dust, not just silicates, may lead to different radiation
pressure. A change in this parameter will affect the drift velocity 
and hence the heating rate and the gas kinetic temperature.
Another factor which affects the CO line fluxes is the adopted
turbulent velocity (Kemper et al. \cite{ciska}). We assume a constant
value of 1 km s$^{-1}$ in our calculation but in reality this could
vary within the circumstellar envelope.
An important issue which we touched upon in the previous section
is the dust drift velocity.
We calculated the drift velocity of $\sim$ 15 km s$^{-1}$ for the averaged
radiation pressure for the grains ($<Q(a,\lambda)>$ = 0.0057).
However, the largest grains with a high drift velocity may no longer be 
momentum-coupled to the gas ($v_{\rm drift} \sim$ 100 km s$^{-1}$!). Hence 
these grains will not partake in the heating of the gas, and hence the 
temperature structure of the envelope should be steeper.
Further progress in this area has to
await a more detailed radiative transfer calculation that includes a
realistic model for the heating and cooling.

\begin{table}
\caption{A comparison of between the observed lines and those from
the models using parameters by Barlow et al. (\cite{barlow}) and
those derived in this paper. We list the available antenna temperatures
(T$_{\rm MB}$), line intensities (I) and line fluxes (F) from various sources.
}
\begin{tabular}{llllll}
\hline \hline
Transition & Observed & Barlow  & Model 1 & Model 2 & Model 3 \\
           &          &  et al. &         &         &         \\
\hline
T$_{\rm MB}$(1-0) & 0.3K    & 4.7     & 0.06    & 0.4     & 0.1  \\
I (1-0)           & 1.5$^{a}$ & 55.0    & 0.8     & 4.6     & 1.57 \\
T$_{\rm MB}$(2-1) & 2.3K    & 20.9    & 0.7     & 5.7     & 1.9 \\
I (2-1)           & 22.0$^{b}$ & 229.6   & 7.5     & 51.4    & 19.3 \\ 
F(16-15)          & 3.3E-20$^{c}$ & 8.0E-20 & 1.6E-20 & 5.9E-20 & 2.9E-20\\
F(17-16)          & 2.3E-20$^{c}$ & 8.6E-20 & 1.7E-20 & 6.2E-20 & 3.1E-20\\
\hline
\label{tab:tmb}
\end{tabular}
\\$^{a}$ CO line intensity (K km s$^{-1}$) from Bujarrabal et al. (\cite{bujar})\\
$^{b}$ CO line intensity (K km s$^{-1}$) Cernicharo et al. (\cite{cerni})\\
$^{c}$ ISO : Barlow et al. (\cite{barlow}) flux in W cm$^{-2}$
\end{table}

Our value of the gas mass loss rate agrees within a factor of two
to three with those from Wannier 
\& Sahai (\cite{wannier}) and Olofsson et al. (\cite{hans}) after scaling 
for the distance and velocity assumed here. However, our value falls short
by an order of magnitude compared with that from Barlow et al.
(\cite{barlow}) who calculated radiative transfer for H$_{2}$O
lines in the LWS wavelength range and from Neufeld et al.
(\cite{neufeld}) who modelled H$_{2}$O lines from the SWS.
The gas mass loss rate derived recently by modelling
H$_{2}$O emission lines of W Hya by Zubko \& Elitzur (\cite{zubko})
is slightly higher than that from Barlow et al. (\cite{barlow}).
In an attempt to reconcile these differences, we can say that our
assumed CO abundance of 3\,10$^{-4}$ may be too high for this
object. But we need to reduce this by a factor of 30
to bring the derived mass loss rate in line with those from the 
water lines. However, the column density of the CO seen in absorption
in SWS seems to contradict the very low abundance (Table~\ref{moltab}).
The abundance of water used in these calculations
may be too low. By increasing the water abundance and lowering the mass loss
rate in those cases, or by increasing our mass loss rate and lowering the
CO abundance in our case, a convergence in mass loss rate can be found 
between these approaches.
Also, it is difficult to get the kinematic information from
ISO LWS observations of H$_{2}$O lines because they are not resolved.
Although SWAS and Odin observations of the 557GHz line should be
resolved, currently the signal-to-noise ratio is not good enough
to be confident about the line profile.
In the future it will be possible to study water lines in more detail
using the heterodyne instrument (HIFI) aboard the Herschel Telescope
due to be launched in 2007 which has a resolution of 1 km s$^{-1}$ or
better.
Equally important, the collisional cross sections between ortho-
and para-H$_{2}$O and ortho- and para-H$_{2}$
for high temperature extending to high quantum numbers will
be available as part of the study project in conjunction with HIFI.
In the case where the lines are subthermally excited, the input
collisional rates have a strong influence in the calculation of
radiative transfer.

\section{Modelling absorption bands}

The near-IR spectrum of W Hya shows several absorption bands due to
various molecules, such as CO, H$_{2}$O. In order to see these bands
clearly, we divided the spectrum by our fit to the SED and normalised
it at 4$\mu$m. We have modelled these absorption bands
assuming thermodynamic equilibrium. The molecular data used are 
taken from Le Floch (\cite{floch}) for the CO molecule, Schwenke (1998, 
private communication) for OH and H$_{2}$O molecules and Hitran 
(Rothman et al., 
\cite{rothman}) for others. The absorption line profiles are
convolved with a Voigt profile with the appropriate full width at
half maximum for direct comparison with the observation
(see Helmich \cite{helmich}). From our SED fit to the spectrum, we
can define the stellar continuum satisfactorily up to
$\lambda \le$ 9$\mu$m.
By fitting the observed continuum-divided
spectrum with our models, we can derive an excitation temperature ($T_{\rm x}$)
for each molecule, along with the column density, $N$, assuming a
Doppler parameter, $b_{D}$. In this paper, we try
fitting the different molecular bands independently and derive the best
set of parameters for each band.
A more sophisticated approach has been made by Cami (\cite{jan})
using a chi-square technique.

\begin{figure}
\resizebox{\hsize}{!}{\includegraphics{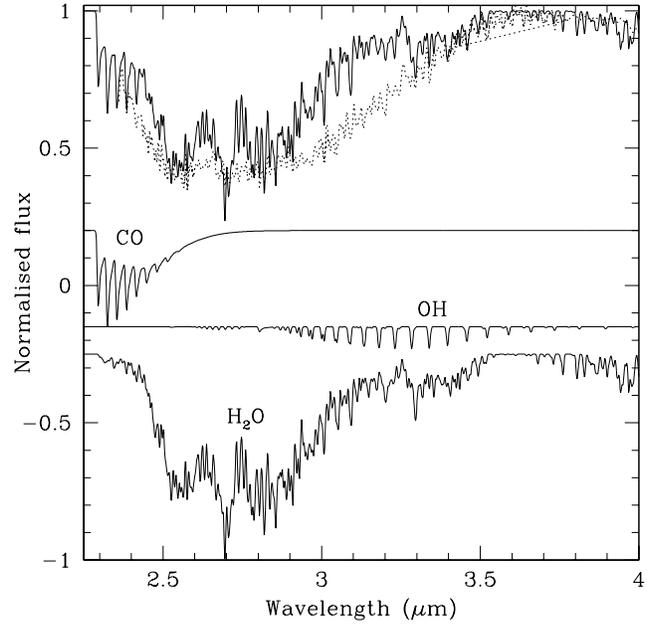}}
\caption{ISO SWS observation in band 1 (dotted line) compared 
with the model absorption of CO $\Delta v$=2, H$_{2}$O
stretching modes and OH absorption bands (solid line).
Each absorption component is also depicted separately.
}
\label{fig:band1}
\end{figure}

\begin{figure}
\resizebox{\hsize}{!}{\includegraphics{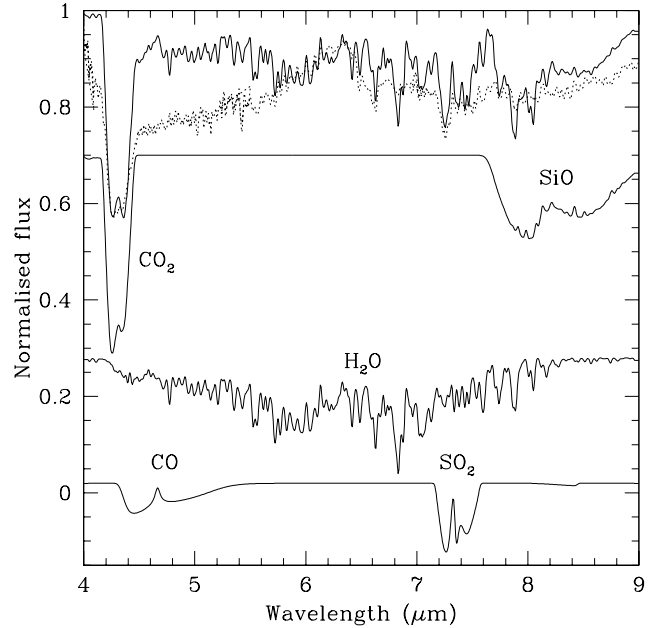}}
\caption{ISO SWS observation in band 2 (dotted line) compared
with the model absorption of CO$_{2}$, H$_{2}$O bending mode,
SO$_{2}$ stretching mode and SiO $v$=1-0 band (solid line).
Each absorption component is individually shown.
}
\label{fig:band2}
\end{figure}

We fitted, in all, six different molecules from 2.3-9$\mu$m. 
The input parameters are listed in Table~\ref{moltab}, while the 
fits are shown in Figs.~\ref{fig:band1} and \ref{fig:band2}. 
From the temperatures, we can
deduce that these molecules originate from different layers in
the circumstellar shell. We see bandheads of 
CO $\Delta v$=2 from $v$=5-3 up to $v$=8-6
which indicates a very hot gas. This is reflected in the input parameter
for CO which has a temperature of 3\,000K. The column density for the
band is high, 10$^{21}$ cm$^{-2}$ also indicating that CO bands originate
from a hot, dense region very close to the stellar photosphere.
We believe that the gas is possibly excited by shocks since the effective
temperature of the star is below the excitation temperature which
gives rise to the observed $v$=8-6 band. This is also consistent
with the inferred Doppler width of 9 km s$^{-1}$ used to model
the CO absorption band.

We also see the stretching modes of H$_{2}$O molecules
at 3$\mu$m. We modelled these bands assuming
an ortho to para ratio of 3. We are able to reproduce the absorption 
band relatively well with a gas temperature of 1\,000K for the H$_{2}$O
stretching modes. However, the band is narrower than
the observed one, especially in the red wing.
The water band overlaps the OH ro-vibrational bands 
($\Pi_{3/2} 1 - \Pi_{3/2} 0$ and $\Pi_{1/2} 1 - \Pi_{1/2} 0$)
which cover from 2.9
to 4$\mu$m. The highly excited lines of OH also indicate that they
come from a warm region, close to the star where H$_{2}$O is
probably photodissociated by shocks.
We also modelled the CO$_{2}$
stretching mode at 4.25$\mu$m. The band seems to come from the same
region as the others. The width of the band is consistent with a
gas of 1\,000K. 

The fit to the bending mode of H$_{2}$O at 6$\mu$m
is, however, not consistent with the observed data. 
In view of the various ratios of adjacent
lines, we estimate that the excitation temperature is only
300K for a column density of 5\,10$^{18}$cm$^{-2}$.
The observed absorption is much deeper than calculated 
in the blue wing where the CO fundamental band is present. 
The 4.6$\mu$m CO R-branch blends with the CO$_{2}$ absorption
wing while its P-branch blends with H$_{2}$O. At the 
resolution of the observations,
individual absorption lines are not resolved. 
We roughly estimate the excitation temperature
of the CO fundamental absorption to be $\sim$ 2\,000K
and put an upper limit for the column density  
of 10$^{19}$ cm$^{-2}$. From very high resolution studies of 
CO lines in Mira variables (Hinkle et al. \cite{hinkle}),
different bands come from different regions, with the fundamental
band coming from a cool region while the overtone bands
come from a hotter region, close to or from the pulsating photosphere.
We can also fit the SiO band at 8$\mu$m with a warm gas of 1\,000K,
possibly from the same layer as the H$_{2}$O stretching modes.
This is consistent with emission coming from gas interior to the dust forming
region where SiO is not bound in dust grains. 
There is evidence
for the weak SO$_{2}$ absorption band at 7.3$\mu$m. 
This band was discovered in a few O-rich AGB stars by
Yamamura et al. (\cite{yama}).
We also see CO$_{2}~\nu_{2}$ absorption band at 15$\mu$m which we
will discuss in more detail in the next section.

\begin{table}
\caption{Parameters for the spectral absorption calculation. For each
molecular band, we also indicate the wavelength coverage.}
\begin{tabular}{llccc}
\hline \hline
 & $\lambda$ & $T_{\rm x}$ & N         & $b_{D}$ \\
 &  ($\mu$m) & (K)      & (cm$^{-2}$) & (km s$^{-1}$) \\
\hline
CO $\Delta v$=2    & 2.3-2.6       & 3\,000 & 1$\,10^{21}$   &  9 \\
H$_{2}$O $\nu_{1}+\nu_{3}$ & 2.3-4.0 & 1\,000 & 5$\,10^{20}$   &  7 \\
OH                 & 2.7-4.0       & 1\,500 & 7$\,10^{17}$   &  8 \\
CO$_{2}~~ \nu_{3}$ & 4.0-4.6       & 1\,000 & 8$\,10^{17}$   &  7 \\
CO $v$=1-0         & 4.2-5.5       & 2\,000 & 5$\,10^{18}$   &  8 \\
H$_{2}$O $\nu_{2}$ & 4.0-9.0       &  300   & 5$\,10^{18}$   &  3 \\
SO$_{2}$           & 7.1-7.7       &  500   & 2$\,10^{17}$   &  3 \\
SiO $v$=1-0        & 7.2-9.0       & 1\,000 & 1$\,10^{19}$   &  7 \\
CO$_{2}~~ \nu_{2}$ & 14.8-15.0     & 1\,000 & 1$\,10^{17}$   &  7 \\
\hline
\label{moltab}
\end{tabular}
\end{table}

Although the overall fit to each band is good, there is still some
residue which we are not currently able to account for. There is
an absorption band just longward of the 4.25$\mu$m CO$_{2}$ band.
It is likely that the hotter component of water from the stretching
modes contributes between 4.5-6$\mu$m. We compared our results to that
from Cami (\cite{jan}; Table 6.2) and found that although our derived excitation
temperatures agree within a factor of two, the column densities
are very different (up to two magnitudes difference) in the cases of 
OH and SiO, while others agree reasonably well. 
This may lie in the fact that we adopt different Doppler
broadening for different regions, i.e., the hotter region has a higher
thermal broadening than the cooler one, while Cami (\cite{jan})
assumed a constant velocity of 3 km s$^{-1}$ throughout. As it turns
out, different combinations of column density and Doppler
width can give similar resulting absorption. The depth of the absorption 
is very sensitive to the adopted Doppler velocity used, more so than the 
column density.

\section{Molecular emission lines}

Both Barlow et al. (\cite{barlow}) and Neufeld et al. (\cite{neufeld})
reported emission due to H$_{2}$O rotational lines from LWS and
SWS FP mode observations, respectively. From the full
SWS scans, we detected 10 more H$_{2}$O rotational lines in the region 
between 30-45$\mu$m. We estimated the line fluxes by assuming a gaussian
profile and these are reported in Table~\ref{tab:water}.
Some lines are a blend of the ortho- and para-water as the wavelengths of
their transitions are too close to be resolved by the SWS scan.
Barlow et al. (\cite{barlow}) 
also noted some CO lines present in their LWS spectrum.
Surprisingly, only two CO lines have been detected in LWS (J=16-15
and J=17-16, see Sect. 3). In this section, we will limit our discussion 
mainly to new results on CO$_{2}$.

\begin{figure}
\resizebox{\hsize}{!}{\includegraphics{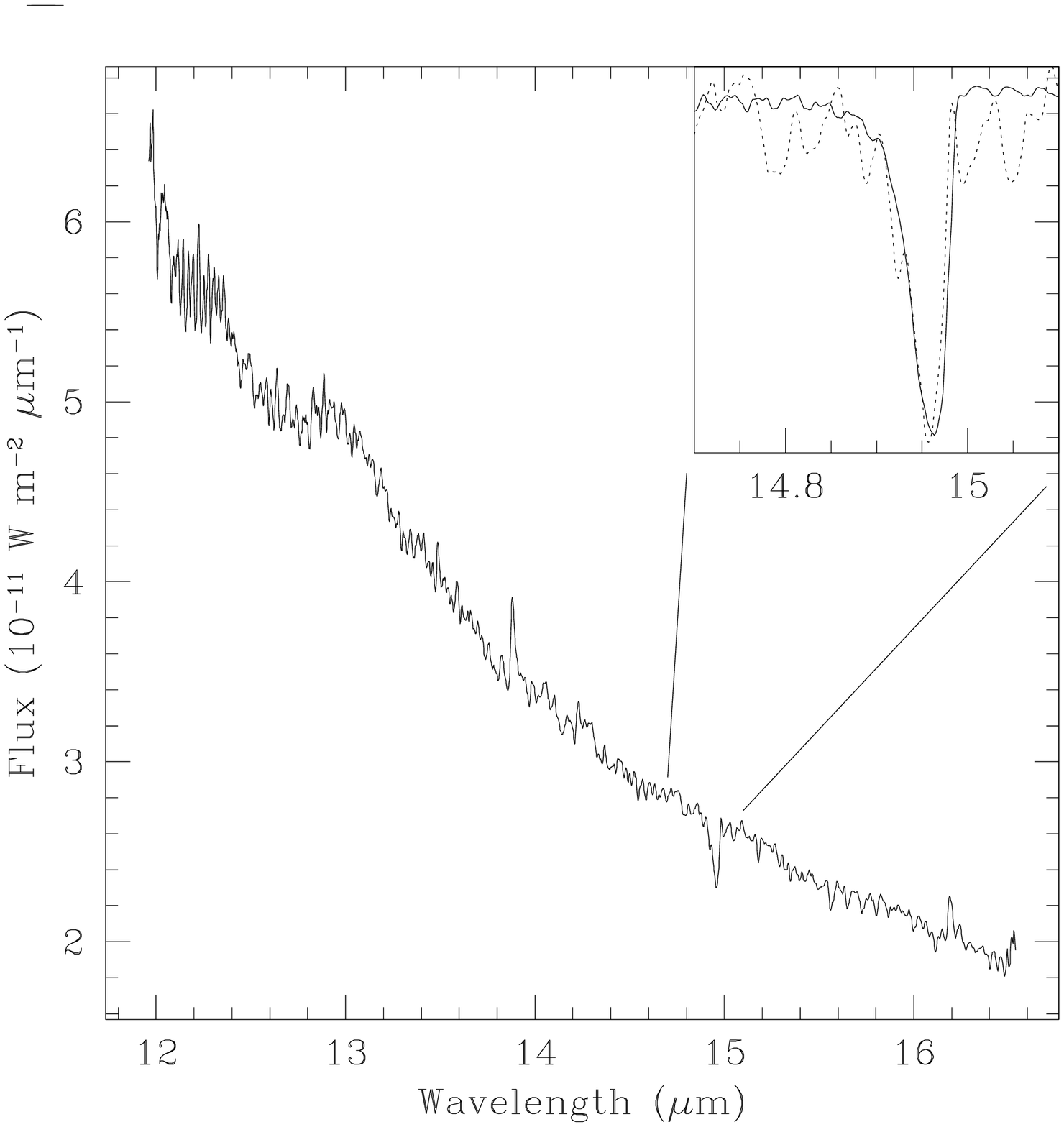}}
\caption{The full grating resolution spectrum of W Hya, showing the
13$\mu$m dust and CO$_{2}$ bands. The insert show the modelled
absorption (solid line) compared to the observed spectrum (dotted line).
}
\label{fig:aot6}
\end{figure}

\begin{table}
\caption{Line fluxes of rotational water lines detected in the ISO-SWS
wavelength range.
}
\begin{tabular}{ccl}
\hline \hline
wavelength  &  line flux       & transition \\
($\mu$m)    &   (W cm$^{-2}$)  & \\
\hline 
32.98 & 1.9E-18 & $6_{61}-5_{50}$, $6_{60}-5_{51}$ \\
35.45 & 5.3E-19 & $5_{33}-4_{04}$, $7_{43}-6_{34}$ \\
35.92 & 4.3E-19 & $6_{52}-5_{41}$, $6_{51}-5_{42}$ \\
39.35 & 1.6E-18 & $5_{50}-4_{41}$, $5_{51}-4_{40}$, $6_{42}-5_{33}$ \\
40.31 & 4.8E-19 & $6_{43}-5_{32}$ \\
40.71 & 1.4E-18 & $4_{32}-3_{03}$, $6_{33}-5_{24}$ \\
43.91 & 6.6E-19 & $5_{41}-4_{32}$ \\
44.14 & 6.1E-19 & $5_{42}-4_{31}$ \\
44.68 & 4.4E-19 & $8_{36}-7_{25}$ \\
45.10 & 6.8E-19 & $5_{23}-4_{14}$ \\
\hline
\label{tab:water}
\end{tabular}
\end{table}

\begin{figure*}[t]
\resizebox{17.5cm}{!}{\includegraphics{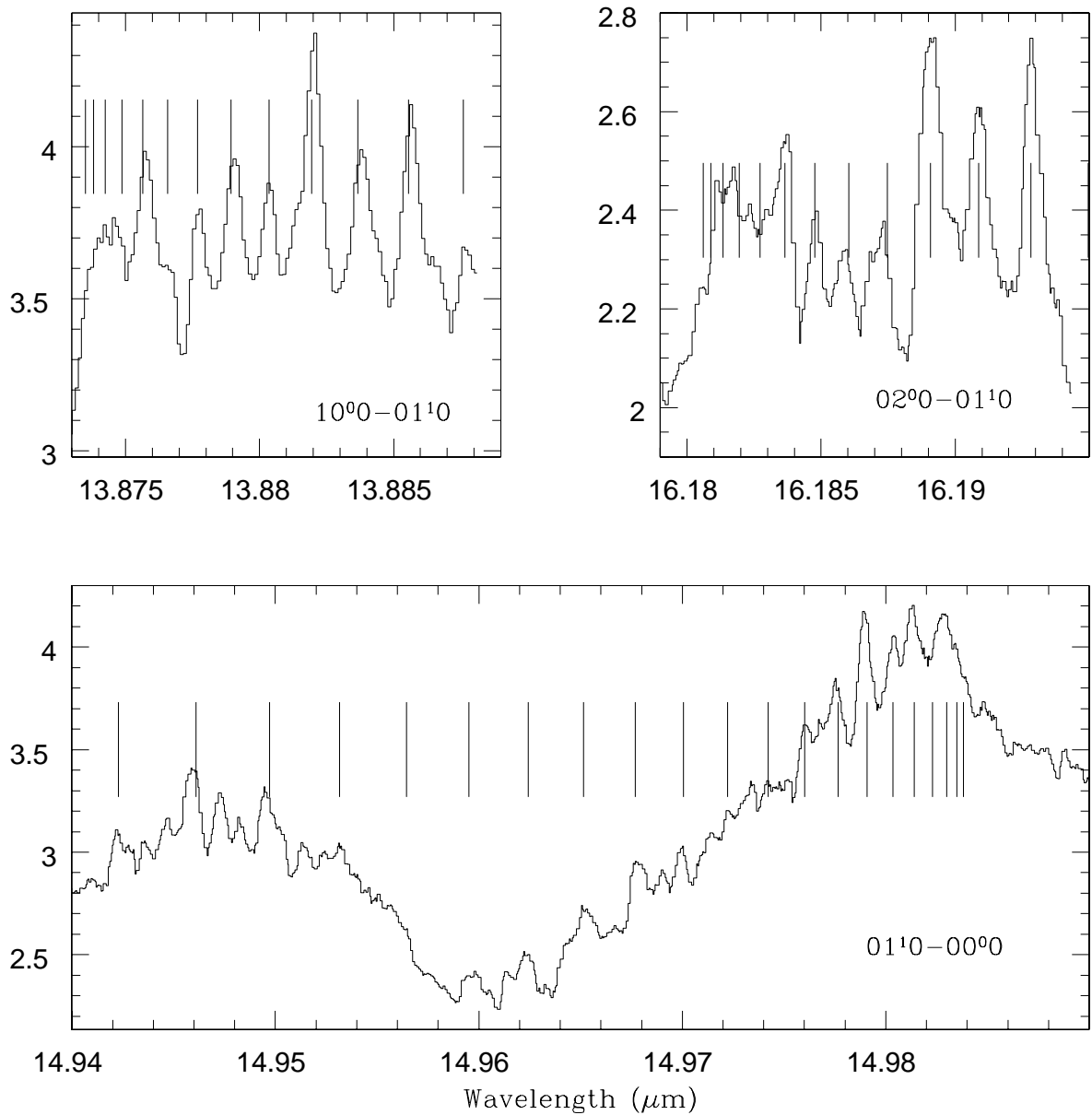}}
\hfill
\parbox[b]{17.5cm}{
\caption{SWS-FP observations of CO$_{2}$ in W Hya with the wavelength
corrected for the heliocentric velocity. The top
panels show the partly resolved Q-branches of the 13.87 and
16.18$\mu$m bands and the bottom panel shows the 14.97$\mu$m.
band. Flux units are 10$^{-11}$ W m$^{-2}$ $\mu$m$^{-1}$. Vertical
lines indicate positions of Q lines. The first few Q lines are 
not resolved.
}
\label{fig:co2}}
\end{figure*}

\subsection{The SWS grating spectrum of CO$_{2}$}

We detected CO$_{2}$ in W Hya between 13-16$\mu$m from
the full scan SWS AOT1. We see these CO$_{2}$ bands in all of 
the stars which exhibit the so called 13$\mu$m dust feature in our
guaranteed time program (PI. de Jong). From this we proposed that
objects with the 13$\mu$m dust have CO$_{2}$ associated with it
(see Justtanont et al. \cite{jus98}, Fig. 1).
On this basis, we obtained further observing time to scan objects
with this dust feature at the highest grating resolution (AOT6).
All the objects observed in the supplementary program confirm
our idea, i.e., all stars with the 13$\mu$m dust show CO$_{2}$
bands. However, the reverse is not always the case, e.g., 
one of the stars in our program, $o$
Ceti, has no 13$\mu$m dust but shows CO$_{2}$ emission and IRC+10011
also shows CO$_{2}$ while the 10$\mu$m silicate is self-absorbed
(Markwick \& Millar \cite{markwick}). 
Cami et al. (\cite{cami}) show a detailed calculation
of CO$_{2}$ emission bands in EP Aqr, one of the stars in our
program. If the association between the 13$\mu$m dust
feature and CO$_{2}$ gas is real then CO$_{2}$ is very prevalent
because Sloan et al. (\cite{sloan}) argue that 
up to 50\% of O-rich semiregular
variables show such a dust feature. 
The formation of CO$_{2}$ is thought to be via CO+OH
(Nercessian et al. \cite{nerc};
Willacy \& Millar \cite{willacy}) where OH is the product of either
photodissociation or of shock destruction of H$_{2}$O molecules. 

In W Hya, the transitions 10$^{0}$0-01$^{1}$0
at 13.87$\mu$m and 02$^{0}$0-01$^{1}$0 at 16.18$\mu$m are in emission
while  01$^{1}$0-00$^{0}$0 at 14.97$\mu$m is in absorption.
We modelled the absorption band which is shifted
from the normal position as seen in emission in other stars. We
attribute this shift towards a shorter wavelength to the importance of
highly excited bands. To excite these bands (e.g.,
02$^{2}$0-01$^{1}$0, 03$^{3}$0-02$^{2}$0, 04$^{4}$0-03$^{3}$0),
we need the gas temperature to be $\sim$ 1\,000K (Fig.~\ref{fig:aot6}). 
These bands have their Q-branches at successively shorter wavelengths
than the 01$^{1}$0-00$^{0}$0 band. By including these bands in our 
calculation, we are able to match the peak absorption at 14.957$\mu$m
as observed.
One thing to note from our
observations of objects with CO$_{2}$ is that only stars with the
14.97$\mu$m absorption show the corresponding 4.25$\mu$m absorption.
Those which show the 14.97$\mu$m in emission lack the stretching mode
counterpart (for the full sample, see Cami (\cite{jan})). 
This probably indicates that the radiative pumping by 
the 4.25$\mu$m band plays a very minor role, if at all, in populating the 
levels which result in the 13-16$\mu$m emission.

\subsection{The Fabry-Perot observation of CO$_{2}$ bands}

We subsequently obtained high resolution
spectra of these CO$_{2}$ bands using the SWS-FP which
resolved the individual Q branches (Fig.~\ref{fig:co2}).
The small dip at 13.877$\mu$m in the band 10$^{0}$0-01$^{1}$0
in this figure is an instrumental effect due to the end of the
scan of FP wavelength setting. The resolved Q-lines are listed in 
Table~\ref{tab:co2} where the wavelength, the energy of the upper level,
the measured line flux and the identification are given.
The flux errors given represent the estimated uncertainties in the 
continuum and do not take into account possible calibration errors so 
that they should be taken as lower limits. 

It is quite remarkable though that we can discern emission of the
Q-branch lines in the depth of the 14.97$\mu$m band while the absorption
blends into a single band. As discussed in Sect. 5.1, this 
absorption is formed in a hot (1\,000K) gas layer in the outer photosphere
and lines may be smeared out by turbulence ($b_{D}$ = 7 km s$^{-1}$,
see Table~\ref{moltab}) and/or optical depth effects.
The first of the individual Q-line to be resolved in emission is Q10.
The close spacing between Q2-Q8 means that even at this resolution
we cannot separate them from each other. We conclude that the 
emission lines are produced in a gas layer which is more extended 
and lying farther out in the circumstellar shell than the hot layer which 
is responsible for the absorption band at 14.95-14.98$\mu$m,
in line with earlier suggestions by Justtanont et al. 
(\cite{jus98}) and Cami et al. (\cite{cami}).

\begin{figure}
\resizebox{\hsize}{!}{\includegraphics{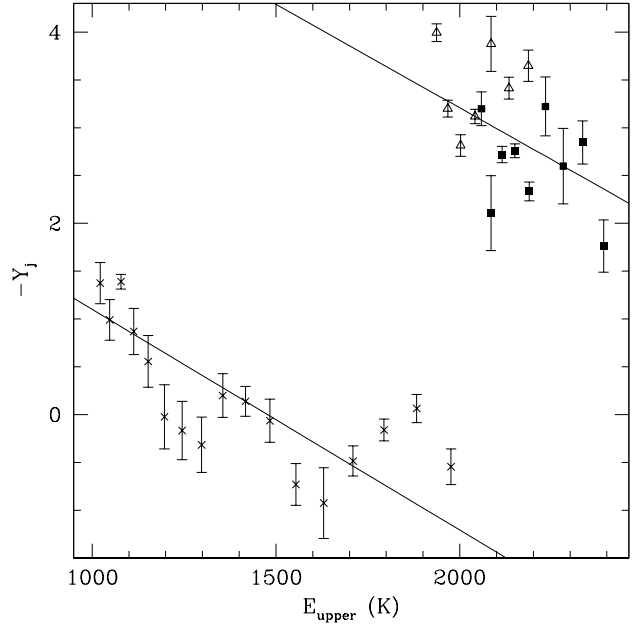}}
\caption{The rotation diagram of CO$_{2}$. Crosses are from
the 01$^{1}$0-00$^{0}$0, triangles represent the 02$^{0}$0-01$^{1}$0 and 
squares represent 10$^{0}$0-01$^{1}$0. The solid lines are the best
fit for these bands.
}
\label{fig:rot_co2}
\end{figure}

\begin{table}
\caption{A list of the CO$_{2}$ emission lines detected from the
FP observations.}
\begin{tabular}{cccl}
\hline\hline
wavelength  & E$_{upper}$      & line flux      & transition \\
($\mu$m)    & (K)      & (10$^{-19}$W cm$^{-2}$)  & \\
\hline
13.87564 & 2059.0606 & 2.6$\pm$0.5 & 10$^{0}$0-01$^{1}$0 Q10 \\
13.87658 & 2084.8831 & 1.0$\pm$0.5 & 10$^{0}$0-01$^{1}$0 Q12 \\
13.87768 & 2115.1954 & 2.2$\pm$0.2 & 10$^{0}$0-01$^{1}$0 Q14 \\
13.87894 & 2149.9974 & 2.7$\pm$0.2 & 10$^{0}$0-01$^{1}$0 Q16 \\
13.88036 & 2189.2882 & 1.9$\pm$0.2 & 10$^{0}$0-01$^{1}$0 Q18 \\
13.88194 & 2233.0677 & 5.3$\pm$1.9 & 10$^{0}$0-01$^{1}$0 Q20 \\
13.88367 & 2281.3351 & 3.1$\pm$1.5 & 10$^{0}$0-01$^{1}$0 Q22 \\
13.88556 & 2334.0896 & 4.3$\pm$1.1 & 10$^{0}$0-01$^{1}$0 Q24 \\
13.88760 & 2391.3305 & 1.6$\pm$0.5 & 10$^{0}$0-01$^{1}$0 Q26 \\
14.94229 & 1976.2415 & 1.6$\pm$0.3 & 01$^{1}$0-00$^{0}$0 Q42 \\
14.94609 & 1882.9064 & 2.8$\pm$0.5 & 01$^{1}$0-00$^{0}$0 Q40 \\
14.94972 & 1794.0594 & 2.1$\pm$0.3 & 01$^{1}$0-00$^{0}$0 Q38 \\
14.95317 & 1709.7018 & 1.5$\pm$0.3 & 01$^{1}$0-00$^{0}$0 Q36 \\
14.95643 & 1629.8354 & 0.9$\pm$0.4 & 01$^{1}$0-00$^{0}$0 Q34 \\
14.95952 & 1554.4611 & 1.0$\pm$0.3 & 01$^{1}$0-00$^{0}$0 Q32 \\
14.96243 & 1483.5803 & 1.9$\pm$0.5 & 01$^{1}$0-00$^{0}$0 Q30 \\
14.96515 & 1417.1943 & 2.1$\pm$0.4 & 01$^{1}$0-00$^{0}$0 Q28 \\
14.96769 & 1355.3043 & 2.1$\pm$0.5 & 01$^{1}$0-00$^{0}$0 Q26 \\
14.97005 & 1297.9109 & 1.2$\pm$0.4 & 01$^{1}$0-00$^{0}$0 Q24 \\
14.97223 & 1245.0155 & 1.2$\pm$0.4 & 01$^{1}$0-00$^{0}$0 Q22 \\
14.97422 & 1196.6187 & 1.3$\pm$0.5 & 01$^{1}$0-00$^{0}$0 Q20 \\
14.97603 & 1152.7216 & 2.1$\pm$0.7 & 01$^{1}$0-00$^{0}$0 Q18 \\
14.97765 & 1113.3246 & 2.6$\pm$0.7 & 01$^{1}$0-00$^{0}$0 Q16 \\
14.97909 & 1078.4285 & 3.8$\pm$0.3 & 01$^{1}$0-00$^{0}$0 Q14 \\
14.98034 & 1048.0339 & 2.2$\pm$0.5 & 01$^{1}$0-00$^{0}$0 Q12 \\
14.98141 & 1022.1414 & 2.7$\pm$0.7 & 01$^{1}$0-00$^{0}$0 Q10 \\
16.18366 & 1937.0742 & 4.2$\pm$0.2 & 02$^{0}$0-01$^{1}$0 Q12 \\
16.18476 & 1967.4081 & 2.2$\pm$0.2 & 02$^{0}$0-01$^{1}$0 Q14 \\
16.18603 & 2002.2345 & 1.7$\pm$0.2 & 02$^{0}$0-01$^{1}$0 Q16 \\
16.18747 & 2041.5522 & 2.5$\pm$0.2 & 02$^{0}$0-01$^{1}$0 Q18 \\
16.18908 & 2085.3610 & 6.0$\pm$2.0 & 02$^{0}$0-01$^{1}$0 Q20 \\
16.19087 & 2133.6598 & 4.1$\pm$0.5 & 02$^{0}$0-01$^{1}$0 Q22 \\
16.19283 & 2186.4477 & 5.6$\pm$0.1 & 02$^{0}$0-01$^{1}$0 Q24 \\
\hline
\label{tab:co2}
\end{tabular}
\end{table}

We have Fabry-Perot observations of individual Q-branch lines for three bands 
of CO$_{2}$ (see Fig.\ref{fig:co2}). In the fundamental band we detected lines 
up to Q42 and in the other bands up to Q24. We can use the observed flux of 
the individual line in each band to construct a rotation diagram from which 
we can derive a rotational excitation temperature T$_{\rm rot}$ characterizing 
the population distribution within the upper vibrational level of each band, 
using the relation
\begin{eqnarray}
E_{j}/kT_{\rm rot} & = & ln(N_{v}/\Phi(T_{\rm rot})d^{2}) -
                   ln(4\pi F_{ji}/A_{ji}h\nu_{ji}g_{j}) \nonumber \\
             & \equiv & ln(N_{v}/\Phi(T_{\rm rot})d^{2}) - Y_{j}
\end{eqnarray}
where $E_{j}$ and $g_{j}$ are the energy and statistical weight of the 
upper state, $F_{ji}$ and $\nu_{ji}$ are the line flux
and frequency of the transition, $A_{ji}$ is the spontaneous
transition probability, $N_{v}$ is the total number of molecules
in the upper vibrational level $v$,
$\Phi(T_{\rm rot})$ is the partition function,
and $d$ is the distance to W Hya. 
Fig.~\ref{fig:rot_co2} shows the rotation diagram of all the resolved
Q-lines. There is considerable scatter in the data points 
due to uncertainties in the calibration (e.g., joining
of different pieces of the spectrum within each band) and in the 
continuum. The offset between the data points in the 01$^{1}$0 level and 
in the 02$^{0}$0 and 10$^{0}$0 levels is due to differences in the number 
of molecules in these levels.

\begin{table}
\caption{Quantities derived from the rotation diagram}
\begin{tabular}{ccccc}
\hline\hline
Level & T$_{\rm rot}$ (K) & $\Phi$(T$_{\rm rot}$) & Y$_{0}$ & $N_{v}$ \\
\hline
10$^{0}$0  &  460  &  4.73  &  -7.57  &  3.0\,10$^{38}$ \\
02$^{0}$0  &  460  &  6.56  &  -7.50  &  4.2\,10$^{38}$ \\
01$^{1}$0  &  435  &  47.2  &  -3.20  &  2.4\,10$^{41}$ \\
\hline
\label{tab:FP}
\end{tabular}
\end{table}

In Table~\ref{tab:FP}, we collect a number of quantities derived from the fits 
in Fig.~\ref{fig:rot_co2}. The derived rotation temperatures are 435 K for the 01$^{1}$0 
level and 460 K for the other two levels combined. The partition functions for each 
vibrational level are computed assuming that the levels are populated in thermal 
equilibrium at the rotational temperature. The quantities Y$_{0}$ are the intercepts 
of the fits in Fig.~\ref{fig:rot_co2} at E$_{j}$ = 0. Putting E$_{j}$ =0 and 
Y$_{j}$ = Y$_{0}$ in Eq. (3) we derive values of N$_{v}$, the total number of 
molecules in each vibrational level, listed in column 5 of Table~\ref{tab:FP}.
These numbers are a lower limit as we assumed that the lines are optically thin.

From these, we then find the 
vibrational excitation temperatures T$_{\rm vib}$ (10$^{0}$0-01$^{1}$0) = 150 K 
and T$_{\rm vib}$ (02$^{0}$0-01$^{1}$0) = 170 K, using the relation 
T$_{\rm vib}$  = 1.44 E$_{ij}$/ln(N$_{j}$/N$_{i}$) with 
E(10$^{0}$0-01$^{1}$0) = 720 cm$^{-1}$ and E(02$^{0}$0-01$^{1}$0) = 617 cm$^{-1}$ 
and taking into account 
that the number of rotational levels in level 01$^{1}$0 is twice as large (J = 
even and odd) as in levels 10$^{0}$0 and 02$^{0}$0 (J = even only). Assuming that 
the relative number of molecules in the ground vibrational state 00$^{0}$0 and 
the first excited vibrational state 01$^{1}$0 is also characterized by 
T$_{\rm vib} \sim$ 160 K we find a total number of CO$_{2}$ molecules of 5\,10$^{43}$.

Clearly the temperature characterizing the distribution of rotational levels in 
the vibrational states of CO$_{2}$ ($\sim$ 450 K) is different from the temperature 
characterizing the distribution of molecules over the vibrational states ($\sim$ 160 K) 
so that the CO$_{2}$ molecules are close to, but not in thermal equilibrium. This 
non-LTE behaviour can be understood on the basis of the symmetry of the CO$_{2}$ 
molecule which does not allow pure rotational radiative transitions and the magnitudes 
of the cross sections for (de-)excitation of CO$_{2}$ by collisions with H$_{2}$, 
which are about three orders of 
magnitude larger for pure rotational transitions than for ro-vibrational 
transitions (Banks \& Clary \cite{banks}). It can be shown by more detailed non-LTE 
multi-level line transfer calculations (de Jong, in preparation) that at a few 
stellar radii from the star and at H$_{2}$ densities above $\sim 10^{8}$ cm$^{-3}$ 
the rotational 
levels are more strongly coupled to the gas kinetic temperature ($\sim$ 400 K) while 
the population of the vibrational levels is dominated by radiation trapping 
in the ro-vibrational lines (the ones that we observe) resulting in a vibrational 
temperature somewhat below the kinetic temperature.

We suggest that CO$_{2}$ absorption and emission originate in two different regions
of an extended dynamical atmosphere :
one close to the star in the outer photosphere where the temperature is 1\,000K,
giving rise to the absorption seen at 4.25 and 14.95$\mu$m, and 
the other farther out at few stellar radii, where the density is still high enough 
to thermally distribute the pure rotational levels at $\sim$ 450 K. Assuming that 
the cooler CO$_{2}$ layer is located at $\sim$ 3  
stellar radii we find, from the total number of molecules observed a column 
density N(CO$_{2}$) $\sim 10^{15}$ 
two  
orders of magnitude smaller 
than that derived for the hot CO$_{2}$ gas (see Table 3) and also several orders of 
magnitude smaller than found for the warm gas in the star EP Aqr
by Cami et al. (\cite{cami}) 

As proposed earlier by Tsuji et al. (\cite{tsuji}), Yamamura et al. (\cite{yama}), 
and Cami et al. (\cite{cami}), 
we believe that the emission line gas detected here is part of the
warm gas layer located between the stellar photosphere and the dust
condensation zone. This gas is probably related to the
blown-up atmosphere observed by Reid \& Menten (\cite{reid}) and found
by H\"{o}fner et al. (\cite{hofner}) in their calculations of the dynamical 
atmospheres of pulsating stars. How CO$_{2}$ molecules are formed in this gas 
is unclear but they may well be a left-over from the preceding 
photosphere phase. According to H\"{o}fner et al. (\cite{hofner}), this gas is 
expected to have 
turbulent velocities of order 1 km s$^{-1}$. Eventually a little 
further out (at $\sim$ 5 stellar radii) the dust is formed which then 
accelerates the flow to 5-10 km s$^{-1}$. The CO$_{2}$ is probably quickly 
photodissociated in the expanding envelope. 
We note here
that the line widths of the emission are much narrower than those of the
absorption, suggesting that the emission comes from a cooler
(since we do not see higher overtone bands in emission), much less turbulent
layer than that which produces the absorption. Whether the hot and cool layers
of CO$_{2}$ are manifestations of the same gas needs to be investigated further
in a detailed model for CO$_{2}$ excitation.


\section{Summary}

By fitting the SED of W Hya from the near- to 
far-IR, we derive a dust mass loss rate of 3\,10$^{-10}$ M$_{\odot}$
yr$^{-1}$ using a combination of three different dust species of
amorphous silicate, Al$_{2}$O$_{3}$ and MgFeO. We also tried adding a
fourth component of spinel to account for the 13$\mu$m dust feature
but in order to fit it, we also expect to see another strong peak at
16$\mu$m which is not present in the observed spectrum. From this, we conclude
that spinel cannot be the carrier of the 13$\mu$m dust emission feature.
Hence, the debate on the carrier of the 13$\mu$m feature is still open.
With the estimated value of the our dust mass loss rate, we get a gas mass 
loss rate range of (3.5-8)\,10$^{-8}$ M$_{\odot}$ yr$^{-1}$ from simple 
consideration of momentum dust driven wind. 
Our derived mass loss rate agrees well with those previously
derived to explain the CO emission, taking into account the differences
in the adopted distance and velocity. However, the discrepancy still
exists between these mass loss rates and those derived from the H$_{2}$O
emission, the latter being consistently higher. The difference
in derived mass loss rates from CO and H$_{2}$O may simply be due to
the abundances assumed for each species.

We obtained the full scan SWS spectrum of W Hya in which we modelled
the absorption bands due to various molecules. The derived excitation
temperatures of different bands suggest that they originate from different
layers of gas, from 300K to 3\,000K. However, some molecules exist
over a wide range of temperatures and cannot really be stratified within
a particular zone, particularly those which show multiple
absorption bands. The exception may be OH which is a product of
water molecules which are dissociated by shocks in the inner part 
of the envelope.

We also observed CO$_{2}$ bands in emission using SWS FP and were 
able to resolve the three strongest Q branches in the mid-IR. The 
resulting rotation diagram yields an excitation temperature of $\sim$450K
while the CO$_{2}$ absorption at 14.97$\mu$m requires a gas of
1\,000K to explain the shift of the peak absorption bluewards. 
The stretching mode at 4.25$\mu$m also requires a layer of
hot gas to explain its width. 
We find from simple calculation based on the rotation diagram
that the vibrationally excited levels may not be in LTE.
Whether the CO$_{2}$ absorption and emission lines can indeed be explained 
by temperature stratification in an extended dynamical atmosphere needs to be 
investigated in a more detailed radiation transfer model.

\begin{acknowledgement}

We are grateful to our referee, Dr. J Alcolea for constructive comments
which improves the discussion in our paper.
This research has made use of the SIMBAD database, operated at CDS
Strasbourg, France.

\end{acknowledgement}


\begin{thebibliography}{}

\bibitem [1989]{anders}  Anders E., \& Grevesse N., 1989, Geochimica et 
                         Cosmochimica Acta 53, 197
\bibitem [1987]{banks}   Banks A.J., \& Clary D.C., 1987, J. Chem. Phys. 86, 802
\bibitem [1996]{barlow}  Barlow M.J., Nguyen-Q-Rieu, Truong-Bach et al.,
                         1996, A\&A 315, L241
\bibitem [1987]{bedijn}  Bedijn P.J., 1987, A\&A 186, 136
\bibitem [1997]{begem}   Begemann B., Dorschner J., Henning T., et al., 1997,
                         A\&A 476, 199
\bibitem [1989]{bujar}   Bujarrabal V., G\'{o}mez-Gonz\'{a}lez J., \&
                         Planesas P., 1989, A\&A 219, 256
\bibitem [2002]{jan}     Cami J., PhD thesis, University of Leiden
\bibitem [2000]{cami}    Cami J., Yamamura I., de Jong T., et al.,
                         2000, A\&A 360, 562
\bibitem [1997]{cerni}   Cernicharo J., Alcolea J., Baudry A., Gonz'{a}lez-Alfonso
                         E., 1997, A\&A 319, 607 
\bibitem [1994]{chapman} Chapman J.M., Sivagnanam P., Cohen R.J., \& Le Squeren
                         A.M., 1994, MNRAS 268, 475
\bibitem [1985]{clark}   Clark F.O., Troland T.H., \& Miller J.S., 1985,
                         ApJ 289, 756
\bibitem [1996]{clegg}   Clegg P.., Ade P.A.R., Armand C., et al., 1996, A\&A 315, L38
\bibitem [1996]{thijs}   de Graauw T., Haser L.N., Beintema D.A., et al., 1996
                         A\&A 315, L49
\bibitem [1984]{draine}  Draine B.T., \& Lee H.M., 1984, ApJ 285, 89
\bibitem [1997]{hip}     ESA, 1997, the Hipparcos and Tycho Catalogues, ESA 
                         SP-1200
\bibitem [1972]{gilman}  Gilman R.C., 1972 ApJ 178, 423
\bibitem [1976]{gold}    Goldreich P., \& Scoville N., 1976, ApJ 205, 144
\bibitem [1994]{groen}   Groenewegen M.A.T., 1994, A\&A 290, 531
\bibitem [1979]{haisch}  Haisch B.M., 1979, A\&A 72, 161
\bibitem [2002]{harwit}  Harwit M., \& Bergin E.A., 2002, ApJ 565, L105
\bibitem [1990]{hawkins} Hawkins G.W., 2000, A\&A 229, L5
\bibitem [1996]{helmich} Helmich F.P., 1996, PhD thesis (Leiden)
\bibitem [1982]{hinkle}  Hinkle K.H., Hall D.N.B., \& Ridgway S.T., 1982
                         ApJ 252, 697
\bibitem [1998]{hofner}  H\"{o}fner S., J$\o$rgensen U.G., Loidl R., Aringer B.,
                         1998, A\&A 340, 497
\bibitem [1998]{jus98}   Justtanont K., Feuchtgruber H., de Jong T.,
                         et al., 1998, A\&A 330, L17
\bibitem [1994]{jus94}   Justtanont K., Skinner C.J., \& Tielens A.G.G.M., 
                         1994, ApJ 435, 852
\bibitem [1992]{jt92}    Justtanont K., \& Tielens A.G.G.M., 1992, ApJ 389, 400
\bibitem [2003]{ciska}   Kemper F., et al., 2003, A\&A, submitted
\bibitem [1996]{kessler} Kessler M.F., Steinz J.A., Anderegg M.E., et al.,
                         1996, A\&A 315, L27
\bibitem [1985]{kn85}    Knapp G.R., \& Morris M., 1985, ApJ 292, 640
\bibitem [1998]{kozasa}  Kozasa T., \& Sogawa H., 1998, Ap\&SS 255, 437
\bibitem [1987]{lane}    Lane A.P., Johnston K.J., Bowers P.F., Spencer J.H., \&
                         Diamond P.J., 1987, ApJ 323, 756
\bibitem [1993]{laor}    Laor A., \& Draine B.T., 1993, ApJ 402, 441
\bibitem [1991]{floch}   Le Floch A., 1991, Mol Phys 72, 133
\bibitem [2000]{leb}     Lebzelter T., Kiss L.L., Hinkle K.H., 2000 A\&A 361, 
                         167
\bibitem [2000]{markwick}Markwick A.J., \& Millar T.J., 2000, A\&A 359, 1162
\bibitem [1977]{mrn}     Mathis J.S., Rumple W., \& Norsieck K.H., 1977, ApJ
                         217, 425
\bibitem [1989]{nerc}    Nercessian E., Guilloteau S., Omont A., \& Benayoun
                         J.J., 1989 A\&A 210, 225
\bibitem [1996]{neufeld} Neufeld D.A., Chen W., Melnick G.J. et al., 1996,
                         A\&A 315, L237
\bibitem [2002]{hans}    Olofsson H., Gonzales-Delgado D., Kerschbaum F.,
                         Sch\"{o}ier F.L., 2002, A\&A 391, 1053
\bibitem [1989]{onaka}   Onaka T., de Jong T., \& Willem F.J., 1989, A\&A
                         218, 169
\bibitem [1999]{posch}   Posch T., Kerschbaum F., Mutschke H., et al., 1999,
                         A\&A 352, 609
\bibitem [1990]{reid}    Reid M.J., Menten K.M., 1990, ApJ 360, L51
\bibitem [1998]{rothman} Rothman L.S., Rinsland C.P., Goldman S.T., et al.,
                         1998, J Quantitative Spectroscopy and Radiative
                         Transfer 68, 665
\bibitem [1999]{ryde}    Ryde N., Eriksson K., \& Gustafsson B., 1999, A\&A
                         341, 579
\bibitem [1982]{schwartz}Schwartz P.R., Zuckerman B., \& Bologna J.M., 1982,
                         ApJ 256, L55
\bibitem [1999]{cjs}     Skinner C.J., Justtanont K., Tielens A.G.G.M., 
                         Betz A.L., \& Boreiko R.T., 1999, MNRAS 302, 296
\bibitem [1996]{sloan}   Sloan G.C., LeVan P.D., \& Little-Merenin I.R., 1996,
                         ApJ 463, 310
\bibitem [2000]{speck}   Speck A.K., Barlow M.J., Sylvester R.J., \& Hofmeister
                         A.M., 2000, A\&AS 146, 437
\bibitem [1979]{spencer} Spencer J.H., Johnston K.J., Moran J.M., et al.,
                         1979, ApJ 230, 449
\bibitem [1990]{stencel} Stencel R.E., Nuth J.A., Little-Marenin I.R., \&
                         Little S.J., 1990, ApJ 350, L45
\bibitem [1998]{szym}    Szymczak M., Cohen R.J., \& Richards A.M.S., 1998,
                         MNRAS 297, 1151
\bibitem [1990]{xander}  Tielens A.G.G.M., 1990, in From Miras to Planetary
                         Nebulae : which path for stellar evolution?, eds,
                         Mennessier M.O., Omomt A., France:Editions Fronti\`{e}res,
                         p. 186
\bibitem [1976]{toon}    Toon O.B., Pollack J.B., \& Khare B.N., 1976, J. Geophys.
                         Res. 81, 5733
\bibitem [1997]{tsuji}   Tsuji T., Ohnaka K., Yamamura I., 1997, A\&A 320, L1
\bibitem [1986]{wannier} Wannier P.G., \& Sahai R., 1986, ApJ 311, 335
\bibitem [1997]{willacy} Willacy K., \& Millar T.J., 1997, A\&A 324, 237
\bibitem [1972]{wilson}  Wilson W.J., Schwartz P.R., Neugebauer G., Harvey P.M., \&
                         Becklin E.E., 1972, ApJ 177, 523
\bibitem [1999]{yama}    Yamamura I., de Jong T., Onaka T., Cami J., \& Waters
                         L.B.F.M., 1999, A\&A 341, L9
\bibitem [2000]{zubko}   Zubko V., \& Elitzur M., 2000, ApJ 554, L137

\end{thebibliography}
\end{document}